\documentclass{aa}  

\usepackage{amssymb} 
\usepackage{multirow}
\usepackage{graphicx}
\usepackage{amsmath}
\usepackage[normalem]{ulem}
\usepackage{booktabs}
\usepackage{tablefootnote} 
\usepackage{txfonts}
\useunder{\uline}{\ul}{}
\usepackage{ulem} 
\usepackage{float}
\usepackage{hyperref}
\usepackage{bm}
\usepackage{threeparttable}
\usepackage{aas_macros}

\begin{document} 

   \title{COMetary dust TAIL Simulator (COMTAILS): A computer code to generate comet dust tail brightness images}

    \author{Fernando Moreno}
          
   \institute{Instituto de Astrof\'{i}sica de Andaluc\'{i}a, CSIC, Glorieta de la Astronom\'{i}a s/n, E-18008 Granada, Spain \\
              \email{fernando@iaa.es}
          }





\titlerunning{COMTAILS, a code to simulate dust tail images}  
 \authorrunning{Fernando Moreno}
 \abstract
   {We present the COMetary dust TAIL Simulator \texttt{(COMTAILS)}, a numerical Monte Carlo code to generate images of dust tail brightness from comets and active asteroids in the Solar System.}
   {We describe a numerical code, available to interested users, capable of generating simulated images of dust tail brightness for comparison with observations, to retrieve key dust parameters including size distribution, ejection velocities, and dust loss rates. An optional stellar field can be included in the background allowing for the assessment of stellar extinction by the tail, which can be compared with observational data.}
   {A Monte Carlo procedure was used to obtain the simulated images. The orbital parameters of the targets and their heliocentric positions and velocities were obtained from the JPL Horizons on-line ephemeris system.}
  {Earlier versions of this code have been used to characterize the dust environment of various targets. In this study we present recent examples that demonstrate its ability to fit observed images of the long-period comets C/2024 N3 (NEOWISE) and C/2023 A3 (Tsuchinshan-ATLAS). The code is readily applicable to future targets that may be identified for the upcoming {European Space Agency} {Comet Interceptor} mission.}
  {}

   \maketitle
%

\section{Introduction} \label{sec:intro}

One of the most powerful techniques for deriving comet particle properties and dust production rates is the interpretation of observed dust tails. In contrast to ion or plasma tails, that result from the ionization of gas molecules by solar ultraviolet light and subsequent de-excitation, forming rectilinear structures pointing away from the Sun, dust tails tend to be much broader and diffuse, generally displaying curved structures. In ground-based telescope images, their appearance depends on the Sun-comet-Earth geometry. As early as 1836, \cite{1836AN.....13..185B} correctly inferred an explanation of comet dust tails that formed as a result of repulsive solar radiation pressure forces. This theory was later improved by Bredichin \citep{1884MmSSI..12..277B}, who first introduced the concepts of syndynes and synchrones ("courbes syndynamiques et synchroniques" in his original French denomination).
It has since become clear that the key parameter involved in dust tail formation is $\beta=1-\mu$, the ratio of solar radiation force to solar gravity force, which depends on particle size and density. A syndyne (or syndyname) represents the locus of particles of a given size (i.e., a single $\beta$ value) emitted at different times, while a synchrone corresponds to the locus of particles of any size (i.e., all values of $\beta$), ejected at a certain time. Both cases assume that the particles leave the nucleus at zero relative speed. It was not until 1968 that \cite{1968ApJ...154..327F, 1968ApJ...154..353F} published their widely recognized studies on the theory of dust tails with application to comet C/1956 R1 Arend-Roland. This is a long-period comet with a prominent antitail dust structure in post-perihelion images. This forward spike was interpreted as dust emission over a specific time interval, beginning approximately two months before the comet's perihelion.  One of the first quantitative approaches to the study of dust tails was given by \cite{1977KimuraLiu}. They explained dust spikes as the collapse of initially spherical dust shells, emitted from the nucleus before perihelion, onto the comet's orbital plane (the second node in their orbits). The authors defined the locus of particles crossing the orbital plane at a given time as the neck-line structure. \cite{1989A&A...217..283F} significantly advanced dust tail modeling by his inverse dust tail technique to retrieve dust loss rates and size distribution functions. This model was applied to numerous short- and long-period comets \citep[see e.g.,][for a review]{2004come.book..565F}  and used a regularization technique 
by \cite{TikhonovArsenin1977} to solve the inverse problem. Variants of this regularization technique have been proposed by \cite{2008Icar..197..183B}, who employed Chebyshev polynomials of selected orders in emission time and particle size, 
and by \citep[]{2009ApJS..183...33M}, where the singular value decomposition method was used. More recently, we approached the problem of dust tail fitting by using a forward Monte Carlo method in conjunction with a multidimensional minimization technique, such as the downhill simplex method 
\citep{1965NelderMead}, to retrieve selected dust parameters. Owing to the high number of free parameters, these methods assume that certain parameters must be estimated by a simple trial and error approach (e.g., the ejection velocities), and cannot ensure that the final fit to the data is unique. These methods currently constitute the most effective approach to this complex problem. 

We are not aware of any published computer code available for use by the cometary and asteroidal communities to generate modeled tails. Therefore, we believe that it would be beneficial to provide the source code for performing this task. In the next section we briefly outline the underlying physics of the problem and describe the procedure used to generate the tail brightness images. The code, written in \texttt{FORTRAN} language, can be retrieved from the web page indicated in Sect. \ref{sec:data_avail}.

\section{General description of the procedure} \label{sec:general}

This section is devoted to the description of the procedure to generate the desired dust tail brightness image. In the current version of the model, the observation point is geocentric,  although this could be switched to another observation point, such as a spacecraft, by modifying the program (specifically, the routine \texttt{SET\_OBSERVATION\_POINT}). The only requirement is that both the target and the observation point must be available in the JPL Horizons on-line ephemeris system  
\footnote{\url{https://ssd.jpl.nasa.gov/horizons/app.html##/}},
 as we used this system to retrieve the position of the observation point and the orbital elements of the target at the observing epoch. The orbital elements were then converted to the heliocentric position and velocity of the target in the orbital plane at the times needed before the observation time. This assumes that the target moves in a Keplerian orbit, and therefore ignores the gravitational perturbations from any other body in the Solar System except the Sun. This approximation is generally valid for tail ages comparable to the target orbital period. For long-term integrations — such as those required for trail computations — it may yield erroneous results if the target has experienced close encounters during the prescribed integration time with one or more massive objects, as its orbit might then significantly deviate from a Keplerian trajectory. This occurs with relative frequency in very long-term integrations of short-period comets. For example, the encounter of comet 67P/Churyumov-Gerasimenko with Jupiter in February 1959 prevents any backward integration before that date, as its orbital elements were significantly different to those of the current epoch.   

The transformation of orbital elements to position and velocity coordinates involves solving the Kepler equations for elliptical and hyperbolic orbits. We accomplished this using the algorithms provided by \cite{1986CeMec..38..307O} and \cite{1988CeMec..44..267G}. The orbital plane coordinates were then transformed into the heliocentric ecliptic reference frame using the standard method \citep[see e.g.,][Sect. 3.5]{1960aitc.book.....S}. 

We set a start and end time for cometary activity, with the end time being the observation epoch.  All times were coded in Julian dates in the main input file. This input file contains all relevant parameters in the simulation. The target itself was identified by its record number that refers to the requested orbit, available at the JPL Horizons on-line ephemeris system.

We worked under the hypothesis that the ejected particles are influenced only by solar gravity and radiation pressure, neglecting all other possible forces, including nucleus gravity or hydrodynamic gas drag. This approximation should be valid beyond $\sim$20 nuclear radii ($R_N$), which we used as the boundary to set the particle terminal velocities of the particles. Given that $\sim 20 R_N$ is a negligible distance compared to the usual target-to-Earth distances, the initial position of the particles at ejection was assumed to be coincident with that of the nucleus position. We further assumed that the particles do not experience any mass loss (e.g., fragmentation) or density change, and that they move in a collisionless regime.

The particles were assumed to be compact spheres of a certain density. This implies that the solar radiation force is strictly radial, resulting in the particles being subjected to the gravity field of the Sun reduced by radiation pressure. Therefore, assuming that they do not experience close encounters with any massive objects in the Solar System, they move in Keplerian orbits around the Sun, which can be elliptical, hyperbolic (attractive or repulsive), or parabolic, depending on the ejection velocity and the $\beta$ parameter. The initial position of the particles at a given ejection time was set coincident with the nucleus position. Particle velocity components were defined relative to the cometocentric frame, based on the selected emission regime described below, and then converted to the heliocentric ecliptic frame. These heliocentric ecliptic velocity components were added to those of the nucleus, resulting in the total heliocentric ecliptic components of the velocity. We converted the initial position and velocity of the particles to their orbital elements, using the transformation described in \cite{1960aitc.book.....S}. From these orbital elements, we then computed the heliocentric ecliptic position of the particles at the observation time. This transformation again involved solving the Kepler equations, using the same algorithms described earlier for the nucleus. The heliocentric ecliptic coordinates of the particle were then converted to the $(L,M,N)$ coordinate system introduced by \cite{1968ApJ...154..327F}. To facilitate a direct comparison with telescopic images, the $(N,M)$ coordinates were rotated to the equatorial coordinate system $(RA,DEC)$ through the target position angle at the observation time,  obtained from the JPL Horizons on-line ephemeris system.

 The brightness associated with each particle depends on its scattering properties. The geometric albedo $p_v(\alpha)$, where $\alpha$ is the phase angle, is related to the $S_{11}(\alpha)$ element of the particle scattering matrix by $p_v(\alpha)=\pi S_{11}(\alpha)/(G k^2)$, where $k=2\pi/\lambda$, $\lambda$ is the effective wavelength of the band filter used, and $G$ is the geometrical cross section of the particle. Since the red wavelengths are the least contaminated by gaseous emissions in comets, in the current version of the code the simulated images correspond to the R-Cousins bandpass. For spherical dust, the scattering matrix element, $S_{11}(\alpha)$, can be computed from Mie theory, given the radius and the refractive index of the particle. However, for low-absorbing particles, the $S_{11}(\alpha)$ is a highly oscillating function of $\alpha$, differing dramatically from those found for equivalent volume non-spherical particles \citep[e.g.,][]{2007JQSRT.106..348M,2021ApJS..256...17M}, such as those found in cometary comae 
\citep[e.g][and references therein]{2021MNRAS.504.4687F}.
To compute the brightness, a more useful approach is to adopt a value of a geometric albedo at zero phase angle. This is combined with an assumed phase function to obtain values of the albedo at observation phase angles different from zero. In the current version of the code, we used the phase function given by David Schleicher  \footnote{https://asteroid.lowell.edu/comet/dustphase/details}, which was constructed from observations of several comets. Future versions of the code will 
benefit from phase function measurements for cosmic dust particles obtained at the {Cosmic Dust Laboratory} (\texttt{CoDuLab}) at the Instituto de Astrof\'\i sica de Andaluc\'\i a  \citep[e.g.,][and references therein]
{2017ApJ...846...85M,2021ApJS..256...17M}. The radiation pressure coefficient and the efficiency for radiation pressure were assumed to take the values $C_{pr}$=1.19$\times$10$^{-3}$ kg m$^{-2}$ \citep[e.g.,][]
{1968ApJ...154..327F} and $Q_{pr}$=1 \citep[appropriate for relatively large absorbing grains, 
see e.g.,][]{1979Icar...40....1B}, respectively.

In the process of building up the synthetic cometary images, we included the brightness of the nucleus, which can be computed as follows. At a given observing time, the following equation holds  \citep[e.g.,][]{1991ASSL..167...19J}:

\begin{equation}
    p_v(\alpha)\pi R_N^2=AU^2 \pi R^2 \Delta^2 10^{0.4(m_\odot - m)},
\label{eq:mag_size}
\end{equation}
where $m$ is the nucleus apparent magnitude, $p_v (\alpha)$ is the geometric albedo, $\alpha$ is the phase angle, $AU$ is the astronomical unit (mean Earth-to-Sun distance, 1 $AU$=1.49598$\times$10$^{11}$ m), $R_N$ is the nucleus radius, expressed in the same units as $AU$, $R$ is the heliocentric distance of the object, $\Delta$ is its geocentric distance (both distances are expressed in astronomical units at the time of observation), and $m_{\odot}$ is the apparent solar magnitude in the corresponding filter. The geometric albedo varies with phase angle according to the phase function. For reference, a typical value of the geometric albedo (at zero phase angle) for comet nuclei is $\sim$0.04 \citep{2019SSRv..215...19F}. The geometric albedo of active asteroids typically ranges from about 0.03 to 0.25, depending on their spectral types \footnote{
https://www.johnstonsarchive.net/astro/astalbedo.html}.
The nucleus brightness extended over the corresponding pixel area, $A$ (expressed in square arcseconds), is given by

\begin{equation}
    S=m+2.5 \log_{10} A,
    \label{eq:surf_br}
\end{equation}

\noindent
which gives $S$ in units of magnitude per square arcsecond.

\noindent
To express $S$ in solar disk intensity units, we used the equation  \citep[e.g.,][]{1992A&A...260..455J}
\begin{equation}
    S=2.5 \log_{10} \Omega + m_\odot -2.5 \log_{10} (i/i_\odot),
\label{eq:solar_units}
\end{equation}

\noindent where $\Omega$ is the solid angle of the Sun at 1 AU,  $\Omega$=$\pi R_\odot^2/AU^2$, $R_\odot$=695660 km is the Sun radius \citep{2008ApJ...675L..53H}, so that 
$\Omega$=6.79$\times$10$^{-5}$, sr=2.890$\times$10$^6$ arscsec$^2$, and $2.5 \log_{10} \Omega =2.5 \log_{10}(2.890\times10^6)$=16.152. The ratio $i/i_\odot$ is the intensity relative to the mean solar disk intensity, with $i_\odot$=F$_\odot$/$\pi$, and F$_\odot$ is the radiation flux density at the solar surface, in the wavelength of interest. 

Algebraic manipulation of Eqs \ref{eq:mag_size}, \ref{eq:surf_br}, and \ref{eq:solar_units}, produced the following expression for the intensity ratio:

\begin{equation}
    i/i_\odot=\frac{p_v(\alpha) R_N^2 \Omega}{AU^2 R^2 \Delta^2 A}.
\label{eq:intensity}    
\end{equation}

The principle of the Monte Carlo procedure is to compute the trajectories of a large number of particles of all sizes within the assumed size distribution, at each time interval within the prescribed ejection time domain, and ejected with a certain velocity function. In the present version of the code, the particles were assumed to be emitted in three different regimes: isotropic emission, sunward emission, and from a selected "active area" on the assumed spherical surface. The actual number of particles ejected as a function of time in a given size bin is a function of the prescribed dust mass loss rate, which depends on the heliocentric distance. In all emission regimes, the ejection of particles is assumed to be uniform, that is, it does not depend on the specific location within the ejection domain.

In the sunward emission regime, particles are ejected exclusively from the illuminated hemisphere of the nucleus, with no emission occurring from the nightside. 

In the active area emission regime, the rotational parameters of the nucleus must first be defined. These include the obliquity (i.e., the inclination of the spin axis relative to the orbital plane, $I$), the argument of the subsolar meridian at perihelion ($\Phi$), and the rotational period. Next, we specified the minimum and maximum values of the latitude and longitude of the active area. As before, particle ejection occurs only from the illuminated portion of the selected active area at each integration time step. Since the dust mass loss rate was prescribed in advance, the initially defined loss rate was adjusted based on the fraction of the active area actually illuminated at each time step. This adjusted rate is provided as an output of the code. If the active area is entirely in darkness, the program skips to the next time step, contributing nothing to the tail brightness.

\begin{figure*}
\includegraphics[angle=0,
trim={1.5cm 2cm 2cm 2cm},clip,width=\textwidth]{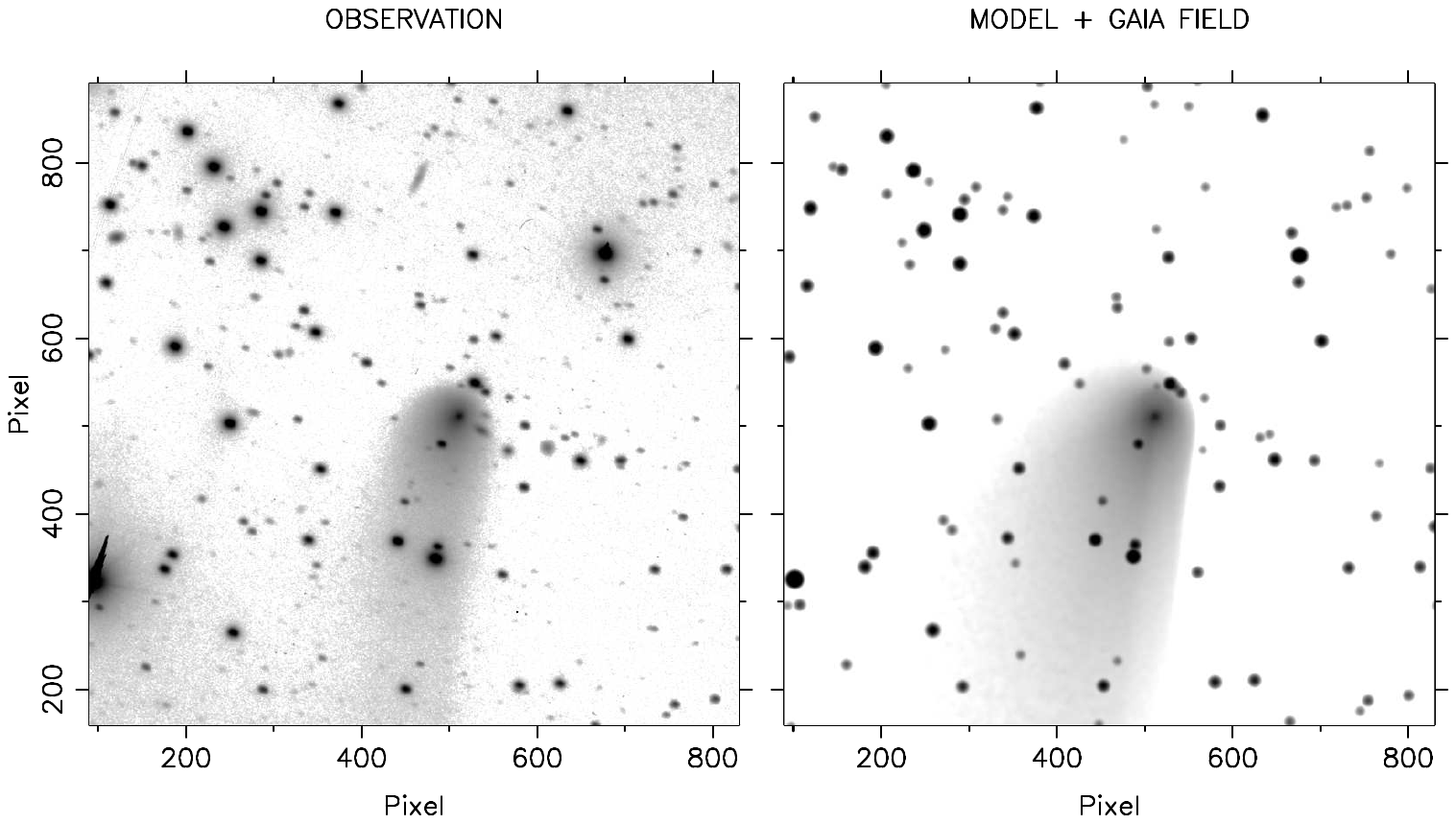}   
\caption{Observation (left panel) and simulation (right panel), using \texttt{COMTAILS}, of comet C/2014 N3 dust tail on December 5, 2015. The depicted image was obtained with the 6-m telescope BTA of the Special Astrophysical Observatory (SAO). Spatial dimensions are approximately 7.4$\times$10$^5$ km, projected onto the sky on a side. North is up and east is to the left.  
\label{fig:2014N3}}
\end{figure*}

Particle speeds were parameterized following common practice, depending on size and heliocentric distance as 
$v(\beta,r_h)=v_0 \beta^\gamma r_h^\Gamma$, where $\gamma$ and $\Gamma$ normally take values of $\gamma$=0.5 and $\Gamma$=--0.5 \citep{1951ApJ...113..464W,2000Icar..148...80R}, and $v_0$ is a constant. In addition, a dependence on the cosine of the solar zenith angle to a certain power can also be incorporated into the velocity expression. The final expression for velocity becomes

\begin{equation}
v(\beta,r_h,z)=v_0 \beta^\gamma r_h^\Gamma \cos{z}^\varepsilon,
\label{eq:velocity}
\end{equation}

\noindent where $z$ is the zenith angle at the emission location, and $\varepsilon$ is an exponent that can  take the value $\varepsilon\sim$0.5 \citep{1997Icar..127..319C}, based on a tridimensional dusty gas-dynamical model.  In addition, a time-dependent velocity coefficient can also be incorporated into the above expression of the velocity, giving a different time dependence to that of $r_h^{-0.5}$.

The number of total simulated particles ejected was given as \texttt{NTIMES x NSIZES x NEVENTS}, where \texttt{NTIMES} in the number of bins in the time domain, \texttt{NSIZES} is the number of size bins within the size distribution defined, and \texttt{NEVENTS} is the number of random directions in space, within the prescribed ejection surface, along which the particles are ejected. A higher number of Monte Carlo events results in an improved signal-to-noise ratio in the dust tail image. The appropriate choice of these inputs depends on the specific problem under investigation. For instance, to generate a trail image, namely, the result of particle ejection during several cometary orbits before the observation date, the \texttt{NTIMES} should be high enough (e.g., several thousand). Otherwise, the simulated image would be too noisy.

The size distribution function was assumed as a differential power-law function, with minimum and maximum radii given by $r_{\text{min}}$ and $r_{\text{max}}$, and power index $\kappa$. The number of particles ejected in a given size bin $N[r_1,r_2]$ and in a given time interval was given by

\begin{equation}
N[r_1,r_2] = A\int_{r_1}^{r_2} r^\kappa \, dr,
\end{equation}

\noindent
where $A$ is a constant. This constant was obtained by considering that the total dust mass ejected in the time interval, $m$, is given by

\begin{equation}
m = \int_{r_{\text{min}}}^{r_{\text{max}}} A r^\kappa \, \frac{4 \pi}{3} \rho r^3 \, dr.
\end{equation}

\noindent
Thus, we had

\begin{equation}
N[r_1, r_2] = \frac{3 m}{4 \pi \rho} \frac
{\int_{r_1}^{r_2}  r^\kappa \, dr} {\int_{r_{\text{min}}}^{r_{\text{max}}}r^{\kappa+3} \, dr}.
\end{equation}

To compute the contribution of the brightness of the ejected particles to a particular pixel on the image, we first calculated the geometric mean, $r_g$, in the interval $[r_1,r_2]$. We verified if this particle falls within the image by using the dynamical equations. The brightness in the corresponding pixel was then incremented by the quantity $(i/i_0)N[r_1,r_2]$, where $(i/i_0)$ is given by Eq. \ref{eq:intensity} for $R_N=r_g$. Since this $r_g$ is taken as representative of the dynamics of all particles ejected in the $[r_1,r_2]$ interval, the number of size intervals must be sufficiently large (i.e., large \texttt{NSIZES}) to ensure an accurate sampling of all particle sizes.

\begin{figure*}
\includegraphics[angle=0,
trim={1cm 2cm 2cm 2cm},clip,
width=\textwidth]{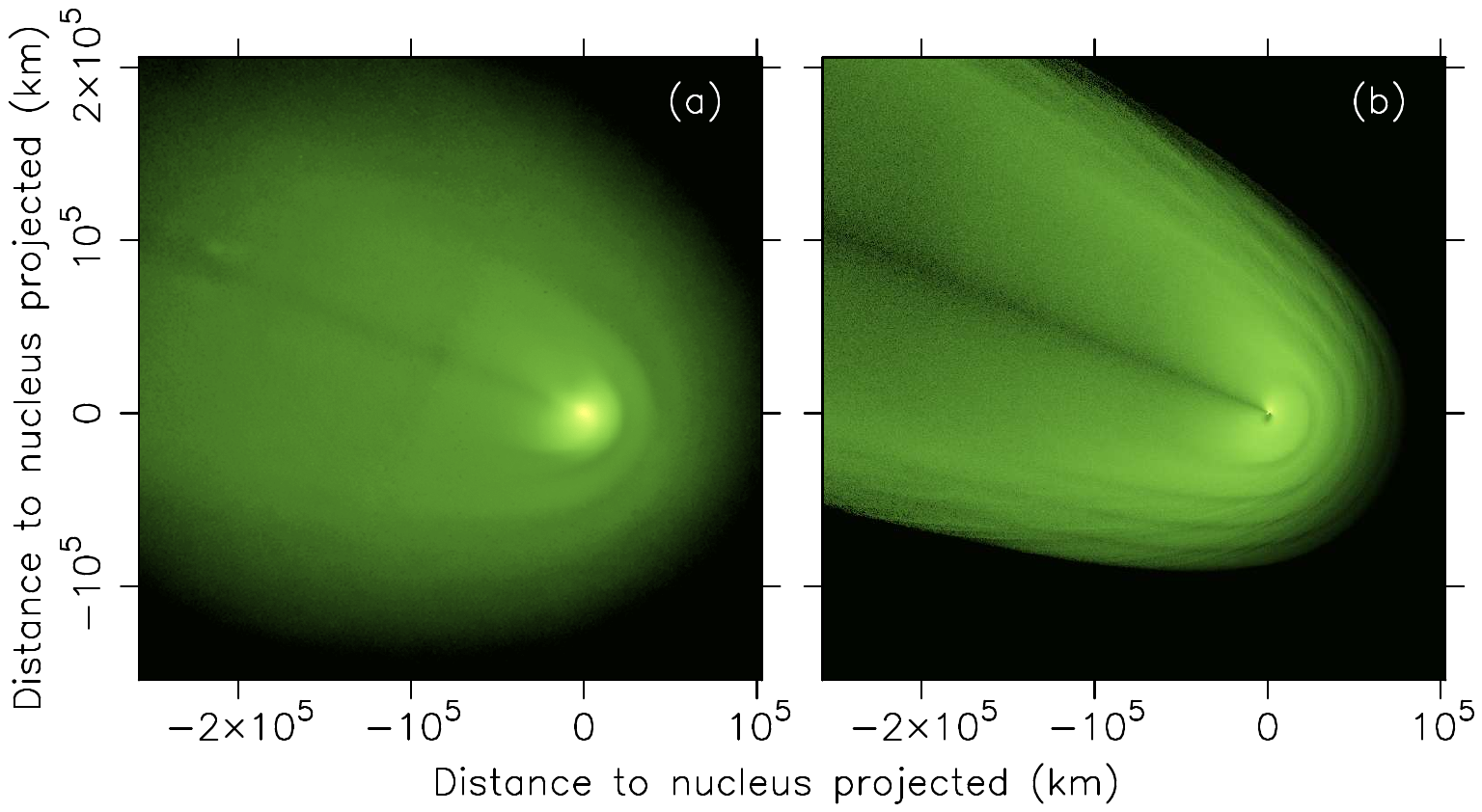}   
\caption{Observation (panel a) and simulation (panel b), using \texttt{COMTAILS}, of comet C/2023 A3 on October 13, 2024.  The depicted observed image was obtained with a Newtonian reflector of 36-cm aperture by Jos\'e Carrillo from the amateur astronomical association  \texttt{Cometas\_Obs}. The simulated image was obtained with \texttt{COMTAILS} assuming an active area anisotropic particle ejection pattern (see text for details). The comet heliocentric and geocentric distances are 0.57 au and 0.47 au, respectively.  North is up and east is to the left.  
\label{fig:2023A3}}
\end{figure*}

\section{Stellar background} \label{sec:stars}

 The images can optionally incorporate the corresponding stellar field from a given catalog on the dust tail image. In the current version of the code, this stellar field is obtained by querying the  \texttt{GAIA} Data Release 3 (\texttt{DR3}) catalog \citep{2022gdr3.reptE...5B} from the IPAC at the California Institute of Technology \footnote{\url{https://www.ipac.caltech.edu/}}. We downloaded the \texttt{GAIA} magnitudes for stellar sources available in the field of view using the $G$, $G_{BP}$, and $G_{RP}$ filters from the catalog. To obtain conventional R-Cousins magnitudes ($R_c$), we used the photometric relationship given by \citep[see][]{2022gdr3.reptE...5B}
\begin{equation}
    y=-0.02275 +0.3961 x -0.1243 x^2 - 0.01396 x^3 + 0.003775 x^4,
\label{eq:phot_conversion}
\end{equation}

\noindent where $x$ and $y$ are the magnitude differences $x=G_{BP}-G_{RP}$ and $y=G-Rc$. 

The stellar magnitudes were then converted to intensity relative to the solar disk using the equation 
\begin{equation}
    i/i_\odot = 10^{0.4(m_\odot+2.5 \log_{10} \Omega -m_\star)},
\end{equation}
where $m_\odot$ and $m_\star$ are the apparent magnitudes of the Sun and the star in the $R_c$ filter, respectively, and $\Omega$ is the solid angle of the Sun at 1 AU, so that $2.5 \log_{10} \Omega$=16.152, as previously shown (see Eq. \ref{eq:solar_units}). The solar apparent magnitude in the $R_c$ filter is taken as $m_{\odot}$=--27.078 \citep{2010AJ....140.1919E}. 

We transformed the ($\alpha,\delta$) equatorial (J2000) coordinates of the stars in the catalog to standard coordinates $(X,Y)$ on the photographic plane, using the standard transformation method

\begin{equation}
\begin{split}    
    X & =\frac{\cos \delta \sin(\alpha-\alpha_0)}{\cos \delta_0 \cos \delta \cos (\alpha-\alpha_0) + \sin \delta\sin \delta_0} \\
    Y &= \frac{\sin \delta_0 \cos \delta \cos (\alpha-\alpha_0) - \cos \delta_0 \sin \delta}{\cos \delta_0 \cos \delta \cos (\alpha-\alpha_0) + \sin \delta\sin \delta_0},
\end{split}    
\end{equation}
where $\alpha_0$ and $\delta_0$ are the equatorial coordinates of the image center. All coordinates are expressed in radians. To obtain the pixel coordinates $(x_p,y_p)$,  each $(X,Y)$ pair was multiplied by the number of pixels per radian characterizing the particular image. We performed a coordinate flip on both axes by setting $x_{pf}=NX-x_p$ and $y_{pf}=NY-y_p$, where $(x_{pf},y_{pf})$ are the requested pixel coordinates on the photographic plane. 

The dust tail partially attenuates the brightness of stellar sources, particularly near the nucleus, where the particle density is usually highest. We calculated the final brightness as 
$(i/i_\odot) e^{-\tau} $, where $\tau$, the optical thickness at each pixel in the simulated image, was computed as the summation extended to all particles falling on the given pixel integrated over time 
\begin{equation}
    \tau(r) = \frac{3 Q_{ext} M}{4 r \rho}. 
\label{eq:opdepth}
\end{equation}
The extinction coefficient is $Q_{ext}$, $M$ is the integrated mass column (kg m$^{-2}$), $r$ is the particle radius, and $\rho$ is the particle density. We assumed $Q_{ext}=2$ (the Fraunhofer approximation) for all particle sizes, although detailed calculations could use Mie theory (in the spherical dust assumption). 

The brightness of the nucleus was reduced by the optical thickness of the coma. However, as the dust particles surround the nucleus, only a fraction of them in the image lie in the line of sight between the observer and the nucleus. The remaining dust particles, being behind the nucleus, do not affect its brightness. We built up the image using the cometocentric coordinate system $(N,M,L)$, introduced by \cite{1968ApJ...154..353F}. Where the $L$ axis is directed from the nucleus toward the observer, those particles attenuating the nucleus brightness must have $L >0$. We used this condition to compute the optical thickness and the resulting attenuation of the nucleus brightness along the line of sight.

To simulate the degradation that affects the observed cometary images owing to the Earth's atmospheric turbulence, the synthetic images can be convolved with a Gaussian function characterized by a specific full width at half maximum. Two \texttt{FITS} files were generated for the resulting brightness images, one written in solar disk intensity units and the other in mag arsec$^{-2}$. In addition, an optical depth image was also written, according to expression \ref{eq:opdepth}. All output images follow the conventional orientation of north being up and east to the left.

\section{Examples of code execution} \label{sec:examples}

The performance of earlier versions of this code has already been demonstrated for numerous comets and active asteroids \citep[e.g.,][and references therein]{2022Univ....8..366M}. This section presents example simulations of unpublished results applied to long-period comets C/2014 N3 (NEOWISE) and the recent Great Comet of 2024, C/2023 A3 (Tsuchinshan-ATLAS). A spectroscopic and photometric study of C/2014 N3 has been conducted and will be published elsewhere \citep{2025Ivanova}. The observed post-perihelion image of C/2014 N3 in Fig. \ref{fig:2014N3} was acquired with the 6-m BTA telescope of the Special Astrophysical Observatory (SAO) on December 5, 2015, when the heliocentric and geocentric distances of the comet were 4.51 au and 3.88 au, respectively. The simulated image, shown in the right panel, assumes sunward particle ejection and includes the GAIA stellar field in the background. Details of the simulation parameters will be provided in the aforementioned publication.

For C/2023 A3, we simulated a stunning image captured on October 13, 2024, by amateur observer José Carrillo from \texttt{Cometas\_Obs} \footnote{\url{www.astrosurf.com/cometas_obs/}}. To reproduce the observed shell structure and the dark linear stripe along the tail axis, we assumed a rotating nucleus characterized by an obliquity of 
$I$=90$^\circ$, an argument of the subsolar meridian at perihelion of $\Phi$=260$^\circ$, and a rotation period of 12 hours, with an active area located between latitudes -45$^\circ$ and 0$^\circ$. The size distribution was dominated by submicron-sized particles. Additional details will be provided in forthcoming publications \citep{Moreno25a,Moreno25b}.


\section{Some technical aspects of \texttt{COMTAILS}}  \label{sec:tech}

As mentioned, the code generates \texttt{FITS} format images, and therefore the \texttt{CFITSIO} subroutine library should be installed on the system. The code also contains the option to graphically display particle positions on the sky plane in the selected spatial domain. This option uses \texttt{PGPLOT} routines, and so this library must also be installed. Internet access is required to retrieve orbital parameters of the target from the JPL Horizons on-line ephemeris system, and to access the \texttt{GAIA} star catalog from the Infrared Processing and Analysis Center (IPAC) at the California Institute of Technology. The code can easily be compiled with the free \texttt{FORTRAN} \texttt{GNU} compiler (\texttt{gfortran}). 

\section*{Data availability} \label{sec:data_avail}

The Monte Carlo code described in this paper is available on {\texttt{GitHub}}\footnote{
\url{https://github.com/FernandoMorenoDanvila/COMTAILS/}}, where the source files and documentation for compilation, execution instructions, and example calculations of the program can be found. 

\begin{acknowledgements}

We acknowledge financial supports from grants PID2021-123370OB-I00, and from the Severo Ochoa grant CEX2021-001131-S funded by MCIU/AEI/ 10.13039/501100011033. 

We are very grateful to the referee for the careful reading of the manuscript and the useful comments.

Jos\'e Carrillo of amateur astronomical association \texttt{Cometas\_Obs} is gratefully acknowledged for sharing the C/2023 A3 image displayed in Fig.\ref{fig:2023A3}, left panel. Oleksandra Ivanova is also gratefully acknowledged for providing the BTA image of C/2014 N3 of Fig. \ref{fig:2014N3}, left panel.

The program \texttt{COMTAILS} makes use of the JPL Horizons 
on-line ephemeris system.

\end{acknowledgements}

\bibliographystyle{aa} 
\bibliography{references} 
\end{document}